\documentclass[review]{elsarticle}

\usepackage{lineno,hyperref}
\modulolinenumbers[5]

\usepackage{algorithmic}
\usepackage{graphicx}
\usepackage{textcomp}
\usepackage{xcolor}
\usepackage{xspace}
\usepackage{multirow}
\usepackage{subcaption}
\usepackage{tabularx}
\usepackage{booktabs}
\usepackage{soul}
\usepackage{url}
\usepackage{longtable}
\usepackage{afterpage}

\usepackage[compact]{titlesec}  

\newcommand{\query}[1]{\textit{`#1'}}
\newcommand{\isq}{QueryQuestions-SE\xspace}

\newcommand{\newtext}[1]{#1}

\newcommand{\updatetext}[1]{#1}

\AtBeginDocument{%
  \providecommand\BibTeX{{%
    \normalfont B\kern-0.5em{\scshape i\kern-0.25em b}\kern-0.8em\TeX}}}

\begin{document}

\begin{frontmatter}

\title{Using Clarification Questions to Improve Software Developers' Web Search}

\author{Mia Mohammad Imran}
\ead{imranm3@vcu.edu}
    
\address{Computer Science\unskip, 
    Virginia Commonwealth University\unskip, Richmond\unskip, Virginia\unskip, USA}

%

\author{Kostadin Damevski}
\ead{kdamevski@vcu.edu}
    
\address{Computer Science\unskip, 
    Virginia Commonwealth University\unskip, Richmond\unskip, Virginia\unskip, USA}
    

\begin{abstract}
{\em Context} 

Recent research indicates that Web queries written by software developers are not very successful in retrieving relevant results, performing measurably worse compared to general purpose Web queries. Most approaches up to this point have addressed this problem with software engineering-specific automated query reformulation techniques, which work without developer involvement but are limited by the content of the original query. In other words, these techniques automatically improve the existing query but can not contribute new, previously unmentioned, concepts.

\noindent
{\em Objective}

In this paper, we propose a technique to guide software developers in manually improving their own Web search queries. We examine a conversational approach that follows unsuccessful queries with a clarification question aimed at eliciting additional query terms, thus providing to the developer a clear dimension along which the query could be improved.

\noindent
{\em Method}

We describe a set of clarification questions derived from a corpus of software developer queries and a neural approach to recommending them for a newly issued query.

\noindent
{\em Results}

Our evaluation indicates that the recommendation technique is accurate, predicting a valid clarification question 80\% of the time and outperforms simple baselines, as well as, state-of-the-art Learning To Rank (LTR) baselines.

\noindent
{\em Conclusion}

As shown in the experimental results, the described approach is capable at recommending appropriate clarification questions to software developers and considered useful by a sample of developers ranging from novices to experienced professionals.
\end{abstract}


\end{frontmatter}


\section{Introduction}

Instead of printed manuals or books, most of software developers' information needs that support their daily tasks are nowadays found through the Web, using a variety of online sources including Q\&A forums, API documentation, tutorials, and chats ~\cite{chatterjee2017what,sadowski2015developers}. Developers regularly search these sites to lookup information, reuse code, and learn new concepts and skills. They usually begin by using Web search engines (e.g., Google, Bing) to locate and access content on popular software engineering sources (e.g., Stack Overflow, W3Schools, Tutorialspoint)~\cite{hora2021googling}. However, numerous developer searches are unsuccessful, consuming valuable developer time and effort. In a recent large-scale study of one million Web search sessions by software developers using Bing, Rao et al. found that software engineering-related queries are less effective than other types of queries~\cite{Rao2020AnalyzingWS}, resulting in higher rates of query reformulations, fewer clicks, and shorter dwell time compared to non software engineering sessions.

Researchers have studied the Web searches conducted by software developers towards understanding key behaviors ~\cite{sadowski2015developers, xia2017developers,hora2021googling}, observing that the primary problem to failed Web searches are poorly constructed queries, i.e., developers commonly fail to specify important technical details in their queries (e.g., specific IDE, operating system)~\cite{rahman2018evaluating}. 
To address the problem of short, incomplete and unspecific queries, researchers have attempted approaches for automated query reformulation that extend a query behind the scenes using synonyms (e.g., using WordNet) or common terms present in highly ranked results~\cite{lu2015query,8530053,9402151,6606630,zhang2017expanding, rahman2019automatic}. However, these automated approaches are limited by the terms in the original query, as they can only act to extend the meaning already present and cannot contribute additional novel context.

To overcome the shortcomings of automated query reformulation, we can aid searchers in manually extending their Web queries by posing clarification questions that directly address a topic that the original query lacks~\cite{aliannejadi2019asking,ren2020conversations}. Clarification questions pinpoint the exact way a developer should extend their query in order to retrieve improved, more relevant results. For instance, recent work by Zhang et al. demonstrates the promise of prompting developers to manually expand their queries in helping them meet their information needs~\cite{zhang2020chatbot4qr}, in their case, for improving question retrieval in Stack Overflow. They proposed conversational query refinement that assists users in recognizing and clarifying the technical details missed in their original queries.
Their approach significantly outperforms other approaches to improve question retrieval in Q\&A sites and helps users recognize missing technical details in their queries through generating clarification questions based on offline analysis of the top-n similar Stack Overflow questions for a query. However, Zhang et al.'s approach is limited to search within Stack Overflow, which is focused only on retrieving questions which are also tagged with keywords (e.g., java, spring). The goal of this paper is to go beyond Stack Overflow to general Web search, which is where most developers begin their information search~\cite{hora2021googling}.

\begin{figure}[t]
\centering
\includegraphics[width=0.9\linewidth]{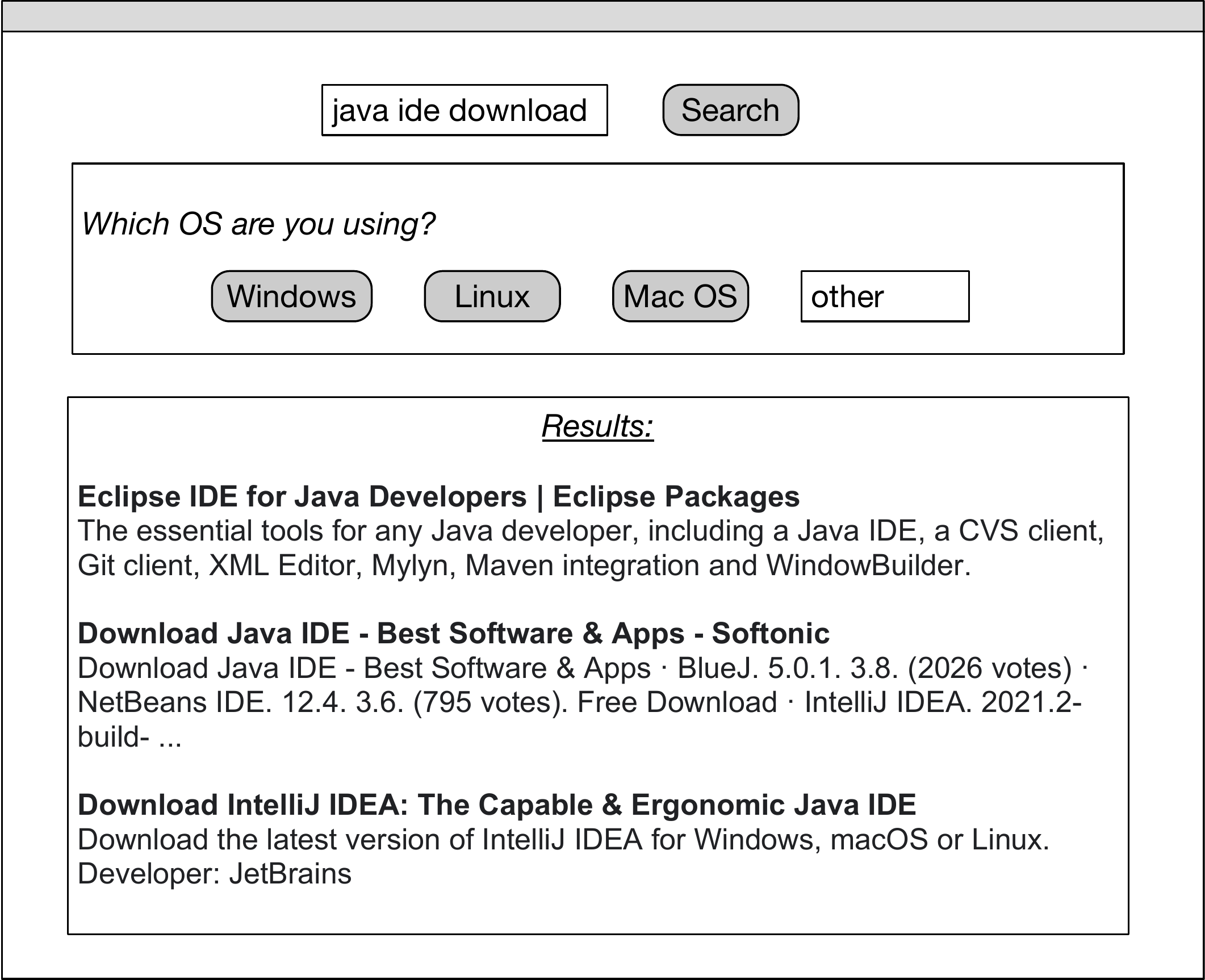}
\caption{An example use of the described system for automatically posing clarification questions to aid developer Web searching.}
\label{fig:motiv}
\end{figure}

More specifically, we envision a scenario where a software developer, e.g., Alice, writes an initial query, which consists of too few terms and gets a set of inadequate results (see Figure~\ref{fig:motiv}). Within the results Web page, our system includes a clarification question that immediately guides Alice in how to expand the query, e.g., {\em Which operating system are you using?}. The system also provides a set of common answers to the question, e.g., {\em Windows, Linux, Mac OS}, for convenience. Alice selects Mac OS and this adds the term to the query and reissues it to the search engine, retrieving another set of results and perhaps posing another clarification question.

In this paper, we analyze a publicly available dataset of Web queries specific to software engineering to devise a set of clarification questions appropriate for this domain. We then devise a neural network based algorithm for automatic recommendation of clarification questions for a developer query that we train with a large dataset created via data augmentation. We evaluate the clarification question recommendation by comparing to a set of baselines. Finally, we build a prototype tool based on the proposed technique as a browser extension that we use to perform a user study to understand both how our technique performs and is perceived by developers as well as the potential of such approaches for manual query extension.

In summary, the key contributions of this paper are:

\begin{itemize}
    \item set of clarification questions appropriate for software engineering query expansion; 
    \item algorithm for recommending clarification questions for a Web search query that is able to handle the small number of terms present in most queries (including a data augmentation step to generate appropriate training data);
    \item quantitative and qualitative evaluation that aids in understanding the value of the technique and the potential of the overall idea.
\end{itemize}

\noindent 
\textbf{ Significance of contributions.} 
This paper is one of the first to propose using clarification questions to improve Web searches conducted specifically by software developers with the goal of improving the quality of retrieved results and, through it, the efficiency with which developers locate relevant information on the Web to aid their daily work. Using clarification questions for general Web search (i.e., so-called conversational search) is recent popular topic in the Information Retrieval community (e.g., see~\cite{aliannejadi2019asking,radlinski17theoretical}).
We contribute a set of clarification questions and a machine learning technique that successfully recommends a relevant question for 4 out of 5 queries. Through a browser extension that implements our technique, we observe that developers value clarification questions as a way to improve their queries and find our technique generally useful. 

A replication package containing the dataset and code used for the study is available~\footnote{ \url{https://anonymous.4open.science/r/Query-Expansion-Questions-5291}}.

\section{Eliciting Clarification Questions}

Software developers' Web queries are often short, consisting of a median of 3 terms~\cite{hora2021googling}, and, therefore, it is common for each query to have multiple {\em facets} (i.e., dimensions of meaning) along which it can be interpreted. Posing clarification questions to developers for an incomplete query aims to identify the specific facet that the developer has in mind, so that it can be specified to the search engine, leading to improved retrieval results. 

With the goal of eliciting a set of clarification questions appropriate for software engineering-related searches, we define an approach that, first,
identifies the facets of a query and, second, uses the facets to create clarification questions that would unravel developer search intent. In this section, we discuss the process we used for eliciting common clarification questions: 1) selecting a corpus of software developer Web queries; 2) identifying the (usually several) facets for each query;  3) generating facet-based clarification questions; and 4) generalizing the facet-based clarification to a set of common clarification questions. 

\subsection{Software Development-Related Queries}

Logs of actual Web search queries are usually not made publicly available by search engines. Rao et al. note that a significant bottleneck in this area of research is the lack of datasets as search logs can not be made public due to privacy laws (see also Section~\ref{related_work})~\cite{Rao2020AnalyzingWS}. In their later work~\cite{rao2020code}, they release a limited dataset of anonymized software development-related search queries mined from Bing logs between September 1, 2019 and August 31, 2020. The dataset contains more than 11,000 real-world search queries related to the C\# and Java programming languages, which they group into 7 categories based on their search intent: API, HowTo, Installation, Debug, Learn, Navigational and Miscellaneous.
They identify that a query is Java/C\# programming language-related when it contains the term \query{java} or \query{c\#} respectively, i.e., each query contains either the term \query{java} or \query{c\#} and at least one other term.

In this work, we select the Java programming language-related queries released by Rao et al.~\cite{rao2020code}, which consists of 6,596 queries. Some of the queries in the dataset are similar (e.g., \query{java api}, and \query{java apis}, \query{java queue}, and \query{java queues}, \query{java for loop}, and \query{for loop java}), and some are noisy (e.g., \query{java chicken}, \query{java apple}). Therefore, we manually filter out queries that are not software development-related or not unique. From the remaining set, we randomly sample a set of 200 queries, ensuring representation from each of Rao et al.'s intent categories (30 queries each from API, HowTo, Installation, Debug, and Learn). The Miscellaneous category was significantly larger than the rest so we sampled 50 queries from it. The category of Navigational queries was excluded as it includes the specific resource or Web page that they developer is navigating to, which makes it inappropriate to our purpose.

\subsection{Facets Identification and Clarification Question Generation} \label{facet-based cq}

A facet is a topic connected to a specific query that describes a dimension of meaning of the query. Facets are often described by a set of semantically related terms that together define a distinct information need or a topic~\cite{wang2009mining, wang2007learn, kong2013extracting, song2011overview, dou2011finding, jiang2016generating}. A typical query usually has multiple, separate facets, e.g., the query \query{java eclipse download} has facets such as \query{Linux},\query{Windows},\query{Mac OS} that describe the operating system used and \query{java 8},\query{java 11} that describe the version of the software.

In order to devise facets for the selected 200 queries, we used four human annotators (including one of the authors of this paper) that are all experienced software developers (2 graduate students, 1 senior undergraduate and 1 industry). Each annotator processed 40-60 queries, based on a detailed set of instructions for generating facets and clarification questions. In order to make sure the annotators were clearly aware of their task, we provided annotation instructions as a written document that the annotators were asked to read to completion before beginning. The instructions also included detailed examples of what to do as well as common mistakes or incorrect approaches to the annotation. We instructed the annotators to use the following two ways of generating facets for each query:

\begin{figure*}[t]
\centering
	\includegraphics[width=1\linewidth]{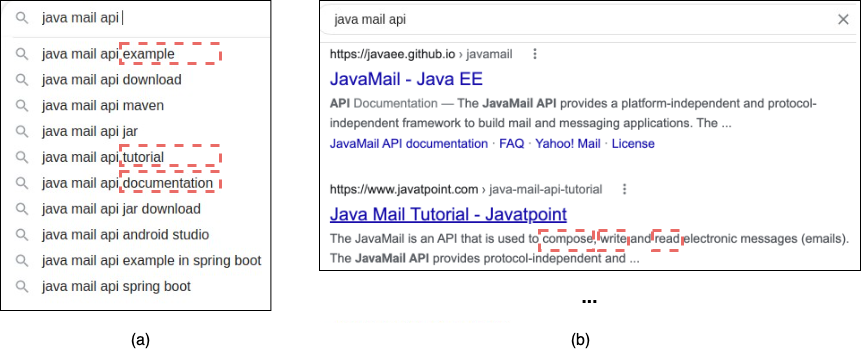}
\caption{(a) Google suggestions for the query - \query{java mail api}, (b) top Google results for the query \query{java mail api}. Using the information in (a) and (b) the annotators identified the facets: 1) `\textit{read}',  `\textit{compose}', `\textit{write}'; 2) `\textit{example}', `\textit{documentation}', `\textit{tutorial}' etc.}
\label{fig:facet-generation}
\end{figure*}

\begin{enumerate}
    \item \textit{Search Engine Top 10 Results}: We asked the annotators to re-enter the query into a popular search engine and examine the top 10 retrieved results in some detail. After examining the results, the annotators were to select the facets that describe the primary characteristics of relevant groups of retrieved results. Sometimes search engine results for the queries included very specific terms that cannot be grouped into facets. For example, for the query, \query{add java to path}, the path - {\em /home/user/path/to/env} - is too specific to belong in a facet, and, for the query, \query{java indexof method} - the method signature - {\em indexOf(String str, int start)} - is too specific. In Figure~\ref{fig:facet-generation}, we provide a screenshot of top results provided by Google and a retrieved top page for the query \query{java mail api}. Here, one possible facet included actions such as `\textit{compose}', `\textit{write}', and `\textit{read}'.

    \item \textit{Search Engine Auto-complete}: Search engines such as Google, Bing, etc., offer query completion suggestions (see Figure~\ref{fig:facet-generation} - left) as the user is typing their query in the search box. These suggested words often describe a key characteristic of a query that can sometimes be appropriate as a facet. We asked the annotators to examine the list of suggested query completions to discover common themes or topics, however, we warned the annotators that sometimes the search engines suggest terms that are too specific to form facets, e.g., (`\textit{example in spring boot}' in part (a) of Figure~\ref{fig:facet-generation}).

\end{enumerate}

The annotators discovered a total of 1,577 facets across the 200 input queries (a median of 8 facets per query).

Next, we asked the annotators to focus on identifying clarification questions for each facet they identified. Specifically, the annotators were instructed to: {\em ``Put yourself in the position of someone who has searched using the provided query. You are to come up with one clarification question that would produce a specific facet.''} 
It is possible to map several facets to a single clarification question and to have multiple clarification question for a single facet. For example, for the query \query{java mail api}, a possible set of facets is:  \textit{1) `example', `documentation', `tutorial', `library'}; 
\textit{2) `maven', `gradle'}; and \textit{3) 
`create', `send'}. Based on these facets, the annotators came up with several clarification questions, such as, `\textit{Are you looking for related API documentation?}', `\textit{Do you want to add a specific library in your build tool?}', and `\textit{What type of mail operation do you want to perform?}'. Finally, the annotators generated a total of 741 clarification questions across 200 queries (a median of 4 clarification questions per query).

\subsection{Grouping into Common Clarification Questions}

The clarification questions generated based on the procedure described above (in Section~\ref{facet-based cq}) are \textit{facet-based clarification questions}. The exact form of these facet-based clarification questions differs from query to query, from facet to facet, and from annotator to annotator. However, it is possible to group the facet-based clarification questions into a set of common, semantically-related clarification questions. For example, the facet-based clarification questions - \textit{`Are you asking about an example?'}, \textit{`Do you want an example of this API?', `Do you want an example of an abstract class or method?'} - can be expressed by the common clarification question - \textit{`What type of document are you interested in (example)?'}. To group the facet-based clarification questions we: 1) group all the similar query-facet pairs; and 2) examine the corresponding facet-based clarification questions to formulate a more generic way of expressing them. 

Two human annotators (from the initial set of four, including one of the authors) generated common clarification questions for all facet-based clarification questions. The annotators worked independently and then held a Zoom meeting to discuss disagreements and agree on appropriate clarification question language. There were a number of facet-based clarification questions where the annotators were not able to devise a general-purpose common clarification question as the information need was very specific to the particular query. For example, for the query \textit{`binary search method java'} the clarification question \textit{`Do you want a solution leveraging recursion?'} is very specific to the particular query and cannot be grouped with other questions in our dataset. In total, we used 592 out of the 741 clarification questions to construct 16 common clarification questions, discarding 149 questions that could not be grouped. 
The common clarification questions, common answers to these questions, and a example query, facet, and facet-based clarification question are listed in Table~\ref{tab:template_questions}. Common answers are the most common facets regarding the clarification questions. For simplicity, from this point in the paper, we will refer to the common clarification questions as \textit{clarification questions}.

\afterpage{%
\begingroup
\small
\setlength\LTleft{-4em}
\begin{longtable}{ p{2em} | p{20em} | p{20em} } 
\hline
\multirow{2}{1em}{}& \multirow{2}{20em}{\textbf{Clarification Question and Common Answers}}
& {\multirow{2}{20em}{\textbf{Example Query, Facet, and Facet-based Clarification Question}}}
\\& \\& \\

\hline
    
CQ1 & 
What type of document are you interested in? \newline
{\em A:} documentation, example, tutorial, use case, performance, books& 
{\em Query:} \query{java reflection api}; 
{\em Facet:} documentation \newline
{\em Q:} Do you want to read API documentation?
    
\\

 \hline
 CQ2 & 
 What type of source code artifact are you interested in? \newline
 {\em A:} class definition, API, framework, library, tool, plugin & 
 {\em Query:} \query{java immutablelist api}; 
 {\em Facet:} Guava \newline
 {\em Q:} Are you referring to Google library Guava?
 
 \\
 
 \hline
      CQ3 & Which IDE are you using? \newline
      {\em A:} IntelliJ, Eclipse, PyCharm, Jupyter, Visual Studio, Xcode
      & {\em Query:} \query{java is not recognized as an internal command}; 
        {\em Facet:} intellij \newline
        {\em Q:} Which IDE are you using?
        \\
 \hline
      CQ4 & What type of operation do you want to perform? \newline 
      {\em A:} read, write, print, parse, override, get, find
      & {\em Query:} \query{java imageio}; 
        {\em Facet:} read, write, resize, show, display \newline
        {\em Q:} What operation on the image do you want to perform?
        \\
 \hline
      CQ5 & What is your file type? \newline
      {\em A:} text, json,	xml, csv, zip, png, jpeg
      & {\em Query:} \query{how to open files with java}; 
        {\em Facet:} zip, jar, rar \newline
        {\em Q:} What's the file type?
        \\
 \hline
      CQ6 & Which system development toolkit are you using? \newline
      {\em A:} Java JDK, iOS SDK, .NET SDK, Android SDK
      & {\em Query:} \query{java was started return code 1}; 
        {\em Facet:} JDK \newline
        {\em Q:} Do you have a JDK installed?
        \\
 \hline
      CQ7 & If you are using a specific tool, framework, or library, which one? \newline
      {\em A:} - 
      & {\em Query:} \query{java.lang.classnotfoundexception:
      com.mysql.jdbc.driver}; 
        {\em Facet:} maven, databricks, gradle, pyspark, spark \newline
        {\em Q:} Are you using maven, databricks, gradle, pyspark, or spark?
        \\
 \hline
      CQ8 & Which version of software are you using? \newline
      {\em A:} - 
      & {\em Query:} \query{install java on raspberry pi}; 
        {\em Facet:} java 8 \newline
        {\em Q:} Are you asking about any specific version?
        \\
 \hline
      CQ9 & What type of installation-related operation are you interested in? \newline
      {\em A:} update, configure, install, uninstall, download, version check
      & {\em Query:} \query{java mongodb}; 
        {\em Facet:} driver \newline
        {\em Q:} Are you trying to download mongodb JDBC driver?
        \\
 \hline
      CQ10 & Which operating system are you using? \newline
      {\em A:} MacOS, Windows, Linux, Android, iOS
      & {\em Query:} \query{java ide download}; 
        {\em Facet:} Windows 10, Mac, Linux \newline
        {\em Q:} Which operating system?
        \\
 \hline
      CQ11 & This seems to be a comparison. Is there a topic you want to compare to? \newline
      {\em A:} - 
      & {\em Query:} \query{kotlin vs java}; 
        {\em Facet:} syntax \newline
        {\em Q:} Do you want to know the syntax difference between Java and Kotlin?
        \\
 \hline
      CQ12 & Which browser are you using? \newline
      {\em A:} Chrome, Firefox, Opera, Safari, Internet Explorer
      & {\em Query:} \query{protection menu allow java};
        {\em Facet:} chrome, firefox, opera\newline
        {\em Q:} Which browser?
        \\
 \hline
      CQ13 & Which data type are you interested in/referring to? \newline 
      {\em A:} integer, string, float, list, map, set, queue
      & {\em Query:} \query{java scanner example};   {\em Facet:} string, char \newline
        {\em Q:} What data type are you trying to scan?
        \\
 \hline
      CQ14 & Are you interested in information related to 32-bit or 64-bit architecture? \newline 
      {\em A:} 32-bit, 64-bit
      & {\em Query:} \query{latest version of java for windows 10};
        {\em Facet:} 32-bit\newline
        {\em Q:} What's the processor version?
        \\
 \hline
      CQ15 & What type of an exception-related operation are you interested in? \newline 
      {\em A:} handle, catch, throw, avoid, implement
      & {\em Query:} \query{missing return statement error java};
        {\em Facet:} exception\newline
        {\em Q:} Are you throwing an exception? \\
 \hline
      CQ16 & What type of debugging-related artifact are you interested in? \newline 
      {\em A:} fix video, fix tutorial, debug, troubleshoot
      & {\em Query:} \query{java package does not exist}; \newline
        {\em Facet:} sub-directory\newline
        {\em Q:} Is the package in right directory? \\
\hline
\caption{List of clarification questions, common answers to each question, and a sample query, corresponding facet, and facet-based clarification questions that were used to generate the specific clarification question.}         

\label{tab:template_questions}
\end{longtable}
\endgroup
}

\section{Recommending Clarification Questions}

In this section, we describe our system that is able to take a user query as input and produce a ranked list of clarification questions according to the their suitability to enhance the query. However, while 200 queries are enough to elicit a reasonable set of clarification questions, they are not adequate to formulate and train a general machine learning model. In addition, to train such a system, we require not just queries, but also their corresponding clarification questions. Therefore, we devise an approach to create more training data via a semi-automatic data augmentation process. Our system (called \isq) for retrieving and recommending clarification questions consists of two parts: 1) data augmentation to generate training data, and 2) neural network architecture to rank clarification questions. 

\subsection{Data Augmentation} In a recent paper, Chen et al. proposed a technique - Local Additivity based Data Augmentation (LADA) - aimed at utilizing a limited set of labeled textual data to generate a diverse augmented dataset in a semi-supervised way~\cite{chen2020local}. LADA can create a large amount of realistic labeled data while improving the generalization of learning. Following LADA's approach, which targets a somewhat different problem - named entity recognition, we design a procedure to generate enough data so that we can effectively train a ML technique for ranking (i.e., recommending) clarification questions based on a software developer query. To augment the data, we leverage RoBERTa model, which is a deep neural model trained on a large-scale corpus of natural language text, to provide prediction of masked words~\cite{liu2019roberta}. The masked words are terms used to augment each query.

More specifically, for each of our 200 real-world queries and clarification questions, we perform the following two augmentation procedures to generate additional queries:

\begin{itemize}
  \item Using RoBERTa's masked word prediction, we \textbf{add} one or two masked terms to the query. That is, we add the \{mask\} term in all gaps between terms of the query and use RoBERTa's suggested replacement for the \{mask\} term. For example, if the original query is \query{java mail api}, for one masked term we create new queries as \query{java \{mask\} mail api}, \query{java mail \{mask\} api}, \query{java mail api \{mask\}}, and \query{\{mask\} java mail api}. Therefore, possible generated queries are \query{java mail api request}, \query{java mail api error}, \query{java mail api documentation}, and so on. We repeat the same process with two masked terms. For each instance of a masked query, we consider the top 100 suggestions generated by RoBERTa.
  
  \item Using RoBERTa masked word suggestion, we \textbf{replace} one or two terms. For example, if the original query is \query{java mail api}, the possible structures of new queries are - \query{java \{mask\} api}, \query{\{mask\} mail api}, \query{java mail \{mask\}}, \query{\{mask\} \{mask\} api}, etc. From these masked queries, possible generated queries are \query{python mail api}, \query{java mail server}, and so on. As before, we consider the top 100 suggestions for each combination.
\end{itemize}

As RoBERTa's suggestions may not be high quality or they may make specific clarification questions that are mapped to a query redundant, we manually verify each suggested query. That is, we validate augmented queries to be \newtext{unique,} software engineering-related, free of noisy terms, and that the clarification questions from the original query are still applicable. 
For example, the annotated clarification questions for the query \query{java mail api} are CQ1 and CQ2. One of RoBERTa suggestions is \query{secure mail api}, which we consider to be of sufficient quality with the clarification questions still applicable. Alternatively, the suggested query \query{java mail api tutorial} is removed because both of the clarification questions are not applicable, i.e., it already specifies the answer to CQ1. 

Our final training dataset consists of a total of 5,151 unique queries and 11,762 clarification questions, a number that includes the original set of input queries. Each query is mapped to one or more valid clarification questions.

\subsection{Neural Network Architecture} \label{nn architecture}
\begin{figure}[t]
	\includegraphics[width=0.99\linewidth]{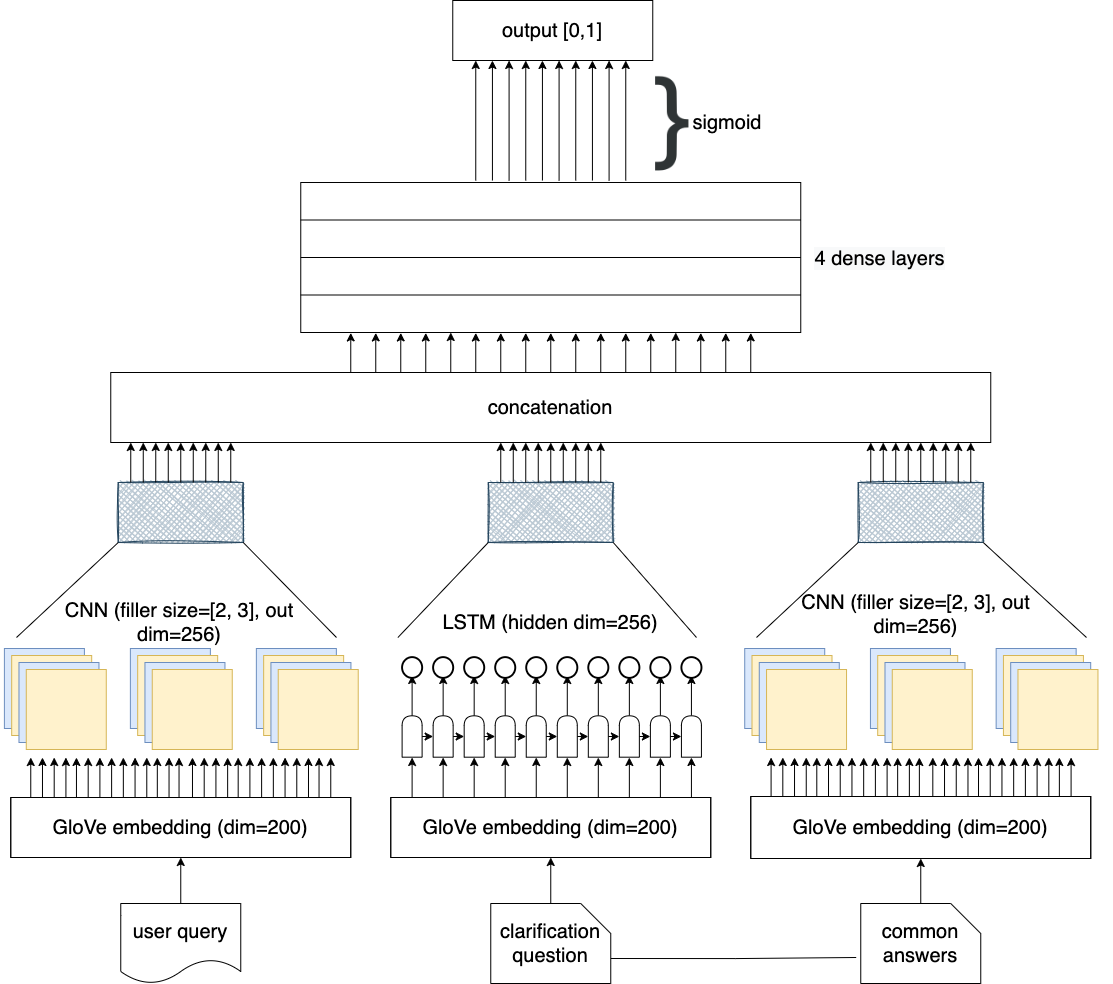}
	\caption{The neural network architecture of the \isq model aimed at determining valid clarification question for software development queries.}
	\label{fig:isq-model}
\end{figure}

Next, we introduce our neural network model for ranking clarification questions. As shown in Figure~\ref{fig:isq-model}, the model computes the probability of a clarification question's validity for a user query using a neural network composed of 3 different parts. Each of the parts centers on each of the 3 inputs in our model: user queries, clarification questions, and the set of common answers. We use a Convolutional Neural Network (CNN) architecture for user queries, a Bidirectional Long Short-Term Memory (LSTM) architecture for clarification questions, and a CNN architecture for common answers. 
A CNN works well for queries and common answers as the order of terms is less important than the different combinations, which a CNN captures well. Clarification questions are natural language sentences so a LSTM would capture this sequence of words well.

For each part of the neural net, as the first layer, in order to introduce additional semantics, we encode the  text with GloVe word embeddings that we pre-trained on the entirety of Stack Overflow (using the Stack Overflow data dump as of June, 2020) with default parameters (vector size = 200; window size = 15). As the second layer, we train the three neural networks separately and merge their outputs. The output of the two CNN layers are two fully connected layers and the output of LSTM layer is a collection of hidden states. When we concatenate the 3 networks, we directly use the CNN's output, while for the LSTM layer we compute the average of the hidden states. As the third layer, we use a dense neural network. We apply the logistic sigmoid function over this output, producing a single floating-point number which can be interpreted as a probability of how much the current clarification question is applicable for the user query. This probability can be also leveraged as a threshold for when not to recommend any clarification questions, i.e., when the query seems clear enough as is (see also Section~\ref{sec:plugin}).

In order to train the model, for each query $q_i$ in the set of all queries $Q$, we create 16 triplets of query, clarification question $CQ_j$, and common answers $CA_j$, one for each of the 16 possible clarification questions. We consider a triplet's label as positive when a clarification question $CQ_j$ is applicable for the query $q_i$ according to the annotation. Note that there is usually more than one clarification question that is appropriate for a single query. We use Binary Cross Entropy loss and the Adam optimizer~\cite{Kingma2015AdamAM}, our implementation leveraging the popular {\sc pytorch} library.

For a newly submitted user query $q$, we use the model for each clarification questions $CQ_j$ and common answers $CA_j$ separately, collecting the resulting output probabilities. Based on the output probabilities $p_{j}$ of each triplet ($q$, $CQ_j$, $CA_j$), we rank the clarification questions from the highest to the lowest probability and recommend the top one or more.

\section{Evaluation}

Our evaluation of the proposed approach can be divided into two parts: quantitative evaluation of the \isq retrieval model and qualitative evaluation via a developer study. The first evaluation strategy aims to find out if clarification questions can be accurately predicted (or recommended) for a newly composed software development-related Web query. The second evaluation strategy focuses on determining if a tool based on the approach described in this paper can be helpful to developers, and what are its strengths and weaknesses. We begin this section by first discussing the quantitative evaluation and then describing the procedure of developer study and collected feedback.

\subsection{Retrieval Evaluation}

\subsubsection{Baselines} As our approach to pose clarification questions is novel to software engineering and there are no similar, comparable techniques that we could compare our model to, we validate the effectiveness of \isq against: 1) {\em Simple Baselines} -- straightforward approaches to ranking that do not require a sophisticated model (e.g., directly computing a similarity based on the GloVe embeddings) and ablation (e.g., using only one part of the model); \updatetext{and 2) {\em Clarification Question Ranking (CQR) Baselines} -- the problem of selecting the highest ranked clarification question from a set of candidate questions is essentially a ranking problem. Therefore, the task of the model is to find the optimal order of candidate questions for a given query. Learning to Rank (LTR)~\cite{liu2009learning} based models rank a set of candidate questions and choose the one with the highest rank, and are a popular approach to compare to when recommending clarification questions~\cite{wang2021template, aliannejadi2019asking, gao2020technical}}.

First, we list the Simple Baselines we identified and used:

\begin{itemize}
    \item {\em {Random}:} We generate a random output of 0 (negative) or 1 (positive) for each combination of query and clarification question.
    
    \item {\em {Similar Embedding ($\delta \geq 0.5$)}:} This baseline generates two vectors: 1) averaged GloVe embedding of the query, and 2) sum of averaged GloVe embedding of the clarification question and averaged GloVe embedding of the common answers. We compute the \textit{cosine similarity} between these two vectors, $\delta$, and set a threshold of $\delta \geq 0.5$ to determine if a query and clarification question are similar.
    
    \item {\em {Similar Embedding ($\delta \geq 0.7$)}:} This baseline is similar to the previous baseline but with a higher similarity threshold of 0.7.
    

    \item {\em {Dissimilar Embedding ($\delta \leq 0.5$)}:} In this baseline, we consider if dissimilarity between the GloVe embeddings of a clarification question (and its common answers) and a query are what determines validity. The intuition is that the more dissimilar a clarification question is the more likely the query is in need to the information it offers. We compute the \textit{cosine similarity} $\delta$ between these two vectors as before and set a threshold of less than or equal to 0.5.
    
    
    \item {\em {Dissimilar Embedding ($\delta \leq 0.3$)}:} This baseline is similar to the previous baseline but with a higher dissimilarity threshold of less than or equal to 0.3.

    \item {\em {Query Only Baseline}:} In this baseline, we use only queries to train the model, ignoring the text of the  clarification question and the common answers. Instead of clarification question text, we use ids that correspond to each clarification questions (1, 2, 3, ..., 16). In the neural network, we use one-hot encoding of the clarification question id.
\end{itemize}

We also used the following set of CQR Baselines.

\begin{itemize}
    \item {\em LambdaRank~\cite{burges2006learning}}: \updatetext{For this baseline, we adopt the pairwise version of the algorithm, where each query and their relevant clarification question (and answers) pair are considered as positive examples. The query and irrelevant clarification questions (and answers) are considered as negative examples. We use the concatenated Glove word embedding (as described in section~\ref{nn architecture}) representation of query, clarification question, and answer used as input features. For LambdaRank implementation, we use Microsoft's LightGBM\footnote{https://github.com/Microsoft/LightGBM} package.}
    
    \item {\em XGBoost ranking~\cite{chen2016xgboost}}: \updatetext{The setup of input documents and input features is as same as LambdaRank. For implementation, we use the open source XGBoost (XGBRanker)\footnote{https://github.com/dmlc/xgboost} package.}
\end{itemize}

\subsubsection{Metrics} We choose several popular information retrieval evaluation metrics, including {\em Mean Reciprocal Rank (MRR)}, {\em Mean Absolute Precision (MAP)}, and {\em Precision@K (P@K)}.

\begin{itemize}
    
    \item {\em Mean Reciprocal Rank (MRR):} The goal of MRR is to evaluate how effective is our technique, or a baseline, in locating the {\em first} valid clarification question. It is computed as: $$MRR = \frac{1}{|Q|} \sum_{i=1}^{|Q|} \frac{1}{rank_{i}}$$, where $Q$ is the set of queries in the test set and ${rank_{i}}$ refers to the rank position of the first relevant clarification question for the $i$-th query.
    

    \item {\em {Mean Average Precision (MAP)}:} For a set of queries, MAP measures how well a model can locate {\em all} clarification questions relevant to a query. MAP is calculated as the mean of average precision values ($AvgP$) for $Q$ queries: $$MAP = \frac{1}{|Q|} \sum_{i=1}^{|Q|} AvgP(i)$$, where $Q$ is the set of queries in the test set and $AvgP(i)$ refers average precision for the query at position $i$.

    \item {\em {Precision@K (P@K)}:} The goal of Precision@K is to measure the number of valid results when considering the top $K$ positions in the ranking. Unlike MRR, it considers all, not only the topmost ranked, results. It is computed as: $$P@K = \frac{1}{|Q|} \sum_{i=1}^{|Q|} \frac{|v|}{K}$$
    , where, as before, $Q$ is the set of queries in the test set and $v$ is the set of valid clarification questions ranked in the top $K$ positions. We use values of 1, 2 and 3 for $K$ as the median of annotated clarification question is 3.
    

\end{itemize}

\begin{table}
\centering
\small
\begin{tabular} {l | c | c | c | c | c }
\hline
    & \textbf{MRR} & \textbf{MAP} & \textbf{P@1} & \textbf{P@2} & \textbf{P@3}  \\ \hline\hline
    {\em{\isq}} & 0.88 & 0.77  & 0.80 & 0.67 & 0.57\\ \hline
    {Simple Baselines:}   &  &  &  &  & \\
    {\quad {Random}} & 0.54 & 0.43  & 0.43 & 0.28 & 0.22 \\
    {\quad {Similar Emb. ($\delta \geq 0.5$)}}  & 0.68 & 0.55 & 0.60 & 0.42 & 0.32 \\
    {\quad {Similar Emb. ($\delta \geq 0.7$)}}  & 0.67 & 0.54 & 0.59 & 0.41 & 0.31  \\
    {\quad {Dissimilar Emb. ($\delta \leq 0.5$)}}  & 0.53 & 0.45 & 0.31 & 0.33 & 0.31  \\
    {\quad {Dissimilar Emb. ($\delta \leq 0.3$)}}  & 0.56 & 0.44 & 0.45 & 0.30 & 0.23  \\
    {\quad {Query Only}}  & 0.56 & 0.48 & 0.41 & 0.31 & 0.27 \\ \hline
    \updatetext{{CQR Baselines:}}   &  &  &  &  & \\ 
    \updatetext{{\quad {LambdaRank}}} & 0.80 & 0.57  & 0.70 & 0.45 & 0.36 \\
    \updatetext{{\quad {XGBoost}}} & 0.79 & 0.56 & 0.71 & 0.45 & 0.35 \\
 \hline
\end{tabular}
\caption{Evaluation results contrasting our system relative to several baselines.}
\label{tab:results}
\end{table}

\subsubsection{Results and Discussion} 
We randomly divide our augmented, annotated dataset using a 80-20 split to evaluate \isq, i.e., 80\% train.

We summarize the results of our technique versus the identified baselines in Table~\ref{tab:results}. The results indicate that our model strongly outperforms all of the baselines, achieving improvements of 0.08 in MRR, 0.2 in MAP, and 0.09 in Precision@1 over the best baseline. The results suggest that the problem of finding a valid recommendation question for a query is not easily solved by the simple baselines we devised. Of the simple baselines, we observe that only the similarity baselines noticeably outperform the random choice; both of the similarity baselines, regardless of threshold, perform nearly the same. The two CQR baselines performed similarly to each other and noticeably outperformed the simple baselines, but were in all metrics weaker than \isq.

From the perspective of whether the \isq model provides usable results, the key metric is Precision@1 as it indicates how well we recommend a single clarification question, which is the most common use case. This metric for \isq is 0.8, which indicates 4 out of 5 times we recommend a valid top most result. \newtext{Precision@2 and Precision@3 decrease significantly from Precision@1 for all techniques including \isq. The reason for this is that a significant subset of queries contain only one valid clarification question and therefore P@1 has much higher ceiling than P@2 and P@3.}

\subsection{Software Developer Study}
To investigate how our \isq model would perform in the real world, we conducted a study of software developers using a prototype of our technique as a browser plugin. We conducted a multi-modal study with a total of 6 participants some of which were novices (i.e., new to software development). First, we collected feedback on the clarification questions from the developers leveraging in-use relevance questions posed after each interaction with the tool. Second, the participants completed a survey after interacting with the tool for at least 2 days. Finally, we discussed their experiences and expectations through an in-person interview. In this subsection, we will first explain the functionality of the browser plugin, then describe the study participants and study procedure, and, finally, report and discuss the collected results.

\begin{table*}[h]
\centering
\small
\caption{Clarification question relevance data collected during use of the plugin.}
\begin{longtable} {p{1.5cm} | p{1.5cm} | p{1.5cm} | p{1.5cm} | p{1.5cm} |  p{2cm} }
\hline
    & 
    \multirow{3}{5em}{\textbf{Industry Exp.}} & 
    \multirow{3}{5em}{\textbf{Active Days}} & \multirow{3}{5em}{\textbf{\# of Queries}}  & 
    \multirow{3}{5em}{\textbf{Relevance}} & \multirow{3}{5em}{\textbf{Usefulness}} \\
    & & & &\\
    & & & &\\
    \hline\hline
      P1 (n) & 0-3 yrs.& 3 & 12 & 10/12 (83\%) & 
      6/8 (75\%), 2 NoAns \\
      
      P2 (n) & 0-3 yrs.& 4 & 23 &14/23 (61\%) & 6/7 (86\%), 7 NoAns \\
      P3 (i) & 4-6 yrs.& 2 & 21 & 15/21 (71\%) & 10/12 (83\%), 3 NoAns \\
      P4 (i) & 4-6 yrs.& 3 & 24 & 18/24 (75\%) & 15/18 (83\%) \\
      P5 (i) & 4-6 yrs.& 4 & 15 & 13/15 (87\%) & 3/4 (75\%), 9 NoAns \\
      P6 (i) & 6-9 yrs.& 6 & 8 & 7/8 (87\%) & 2/6 (33\%), 1 NoAns \\
 \hline
\end{longtable}
\caption{Clarification question relevance data collected during use of the plugin. n = novice, i = industry.}
\label{tab:user_studies}
\end{table*}

\subsubsection{Browser Plugin}\label{sec:plugin} 
\begin{figure}[tb]
    \centering
	\includegraphics[width=0.6\linewidth]{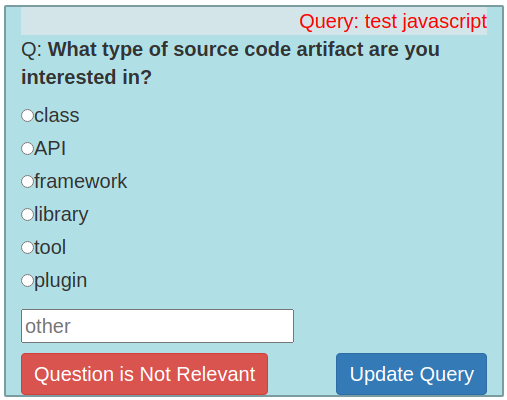}
	\caption{Browser plugin for user study.}
	\label{fig:plugin-user-study}
\end{figure}

To allow study participants to experience receiving clarification questions first hand, we created a Google Chrome browser plugin that monitors user Google queries and reacts with a pop-up if a clarification question is recommended by \isq. If the prediction score is less than 0.5 for all clarification questions, we do not provide any clarification questions as, in this case, it is likely that the developer has an adequate query that does not need clarifying. The pop-up window appears alongside the Google results page and contains: 1) the highest ranked clarification question; 2) a list of common answers and an input box to provide a typed answer; and 3) two buttons (\textit{`Update Query'} and \textit{`Question is Not Relevant'}) at the bottom (see Figure~\ref{fig:plugin-user-study}). When users press the \textit{`Update Query'} button, the plugin appends the answer to the original query and performs a Google search, returning a new set of Google search results on the same page. The user can also click on the \textit{`Question is Not Relevant'} button, which is relevance feedback that we collect for the study.  After selecting \textit{`Update Query'}, alongside the updated Google results page, we ask the user for in-use feedback on the effectiveness of the clarification question via another pop-up that asks \textit{`Was the prior question useful in suggesting how to clarify the query?'} where the user can select either \textit{`Yes'} or \textit{`No'}. The user can also choose to select no option here at all and close the plugin.

\subsubsection{Participants and Procedure}

Our goal was to attract two groups of developers for our study: experienced software developers from industry and novice software developers, with the aim of having representatives from both groups. More specifically, through personal contacts, we recruited 4 software developers currently employed in industry (with industry experience ranging from 4 to 9 years) and 2 novice programmers (students that have recently, in the last 1-2 years, learned how to program) that were currently programming on a daily basis. 

Through a recruitment email, we communicated the basic context of our study and provided a short video on how to install and use the plugin. We asked the participants to watch the video carefully, and then use our plugin during their regular work for software development-related Web searching. 

We asked the participants to let us know once they have used the plugin sufficiently to develop an opinion and feel ready to proceed (after at least 2 days), allowing them to take additional time using the tool if they chose. After the participants contacted us that they are ready, we provided them with a link to a short online survey asking several questions regarding their experiences, including:

\begin{itemize}
    \item How often they used the plugin -- \textit{``How often did you use the plugin in your Google searches?''}, with 3 possible answer options -- {\em `On all searches where I didn't like the initial Google results'}, {\em `On some searches where I didn't like the initial Google results'}, and {\em `On every software development-related Google search'}. 
    \item Clarification question relevance -- \textit{``How relevant were the questions provided by the plugin to your searches?''}, answered by a 5 point linear scale (Least Relevant to Most Relevant).
    \item Clarification question usefulness -- \textit{``How useful were the questions provided by the plugin to finding the information you were looking for?''}, answered by a 5 point linear scale (Least Useful to Most Useful).
    \item Open-ended questions on ideas for improvement -- \textit{``How would you improve the clarification questions asked by the plugin?''}, and \textit{``How would you improve the plugin (apart from the questions themselves)?''}.
\end{itemize}

Following the completion of the study, we organized an online meeting with each participant separately so that they can share their opinions through a semi-structured interview. Our questions in the interview can be organized into 3 topics: 

\begin{itemize}
    \item Questions regarding common search behavior and motivation -- \textit{``Do you write queries that are not specific enough?''}, \textit{``When you search, are you usually satisfied with the results you get?''}.
    \item Questions regarding the clarification questions -- \textit{``Do you think asking clarification question on search queries can be useful?''}, \textit{``How did you find the questions the tool posed?''}, \textit{``How do you think the clarification questions can be improved?''}, \textit{``Can you think of any additional clarification questions?''}.
    \item Questions regarding the plugin --  \textit{``Did you find our plugin useful?''}, \textit{``Did the plugin help you get higher quality results from your Google searching?''}, \textit{``What did you like and dislike about the plugin?''}.
\end{itemize}


\subsubsection{Results and Discussion}

One of the study participants (P1) was not able to schedule the interview for personal reasons. Otherwise, all other participants completed all aspects of the study, \newtext{including the in-use (popup-based) feedback and the post-activity survey}.

\noindent {\bf In-Use Feedback.}
Table~\ref{tab:user_studies} lists the number of queries performed by individual software developers, the number of days the plugin was actively used (days with at least one query performed are counted), and developer feedback on query relevance (as reported by the button on Figure~\ref{fig:plugin-user-study}) and usefulness captured through an additional popup after the query was reformulated \newtext{(and the updated Google results shown to the developer)}. The six developers performed a total of 103 queries, and, of these queries, they deemed 77 (75\%) clarification questions as relevant. Out of these 77 queries, 42 clarification questions are selected as useful, 13 are selected as not useful, and no feedback was provided for the remaining 22 queries; the usefulness rate out of the relevant clarification questions was 76\% (42/55), and 52\% (42/81) overall, not counting the clarification questions without feedback. 

\noindent {\bf Survey.}
On the question - \textit{``How often did you use the plugin in your Google searches?''} - 3/6 developers selected {\em `On all searches where I didn't like the initial Google results'}, 2/6 developers selected {\em `On some searches where I didn't like the initial Google results'}, and 1/6 developer selected {\em `On every software development-related Google search'}.

\begin{figure}[t]
	\includegraphics[width=0.99\linewidth]{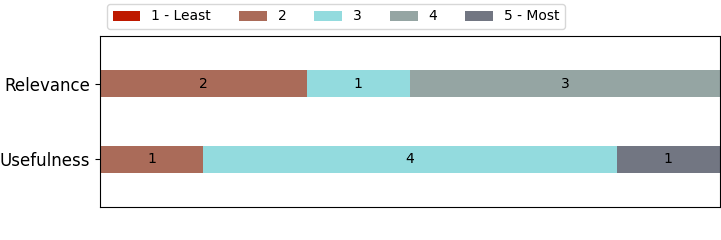}
	\caption{Reported relevance and usefulness of \isq model.}
	\label{fig:viz_relevancy_usefulness}
\end{figure}

The relevance and usefulness of the \isq model as reported by the 6 developers on the survey are shown in Figure~\ref{fig:viz_relevancy_usefulness}. The results indicate that 4/6 developers found the clarification questions to be relevant (with a score of least 3 on the linear scale), and 5/6 developers found the clarification questions useful (with a score of least 3 on the linear scale).

The responses to the two open-ended questions - \textit{``How would you improve the clarification questions asked by the plugin?''}, and \textit{``How would you improve the plugin (apart from the questions themselves)?''} - were repeated with more context in the one-on-one interview, therefore, we discuss them there.

\noindent {\bf Interview.}
We conducted each interview via Zoom. We started each Zoom meeting with a warm-up question asking about the type of software development-related Web searches performed by each participant. Most of the developers identified more than one search scenario, including both exploratory and lookup~\cite{marchionini2006exploratory}. Next, on the question regarding whether they were satisfied with typical search engine results, most participants indicated that they were, however, that they could be improved. When we asked whether clarification questions could be useful in improving search, all of them agreed that this is a good mechanism. Specifically, P4 found the experience of our plugin helpful in this regard, stating \textit{``From my experience in using this plugin, I found clarification questions useful.''} P2 and P5 also mentioned that adding terms to queries using clarification questions is a good idea, while P3 pointed out that clarification questions could help to narrow down the search results and provide more direction to generic search queries. 

As for the effectiveness of the specific clarification questions provided by our plugin, P2 noted, \textit{``they were good as long as they were relevant.''} P3, P4, and P5 also provided similar opinions, however, P6 disagreed. P6 was concerned that our approach would not help a domain-expert experienced developer and in P6's opinion, \textit{``I already knew what I was looking for and search it effectively on Google.''} However, others found the clarification questions helpful, e.g., P3 noted, \textit{``It helped me to clarify contexts on some queries when google results were too broad.''} P2 remarked, \textit{``I liked it when I just didn't know what to add to a search. The tool provided some suggestions.''} P5 noted, \textit{``It did. I was able to get a better query. And because of that better results in the search.''}


The participants offered several ideas to improve the plugin. P4 remarked, \textit{``What I like is that it is easy to use and questions were most of the cases relevant... It may be a good improvement if this can be made more interactive.''} The interactive idea was supported by others as well, e.g., P6 stated \textit{``What kind of queries developers make mostly? - for debugging issues by searching for the specific errors, exceptions, etc; tutorials, examples, documentation of specific library, framework, etc. The clarification question can ask for the type of query first, and based on that it can suggest specific terms to improve the query.''} 

P5 suggested improving the approach by using dynamic options for the common answers suggested by the tool, which would provide common answers that are more tightly integrated to the specific query. P3 and P6 also expressed that dynamically generated answer options could be a way to improve the existing tool. P1 supported this idea as well in her response to the open-ended survey questions. P5 had another suggestion on improving the tool by reformulating the queries multiple ways, e.g., other than adding a term at the end, we could add a term at the beginning, remove unnecessary terms (such as adverbs, pronouns), etc. 

Lastly we asked, {\em ``based on your experience using this plugin [...] would you use it in your daily development work?''} All of the participants except for P6 replied positively. P2 said, \textit{``Yeah, I will. I don't think I would use it like every search. But I think I would use it a fair amount of time.''} P5 remarked, \textit{``Yes, yes, of course, it was helping me get better queries.''} P3 mentioned, \textit{``Yeah, I will use it as the functionality of the tool narrows down the search domain and saves development time.''} P4 commented, \textit{``Why not if the questions are relevant and the tool can serve my purpose?''} P6, however, didn't dismiss using improved version of the tool, \textit{``Maybe (I will) when you could improve the tool as the suggestions we have discussed... I like the approach that the tool is trying to help to search and find better answers.''}


\noindent {\bf Summary.}
The developer study data collected from the 6 software developers indicated that the idea of clarification questions has potential. The clarification questions were mostly valuable (in-use relevance between 61\%-87\% and usefulness between 75\%-86\%, without P6 (33\%); survey at or above 3 out of 5 for relevance in 4/6 participants and for usefulness in 5/6 participants). In one-on-one interviews, they noted that clarification questions can provide convenient suggestions, narrow down search space, and save development time. The key suggestion for improvement was to add more dynamicity and interactivity, e.g., suggesting more bespoke answers to the questions or engaging the developers in a conversation as in conversational search~\cite{radlinski17theoretical}.


\noindent{\bf Comparison to Google's Related Searches.}
\newtext{In order to understand if the clarification questions can provide the developers with a functionality different from what is currently available by popular search engines, we compared the  queries expanded by the developers (using our clarification questions) with the set of query suggestions provided by Google's Related Searches functionality. Google's Related Searches appear at the bottom of each results page, and consist of 8 suggested queries that reformulate the current query provided to the search engine. While the algorithm for selecting the 8 suggested queries is not publicly available, Google states that it is based on a corpus of past search queries.\\
Out of the 42 queries that were successfully expanded by the developers in our study, 9 (21\%) had an exact or near exact match among the queries suggested by Google Related Searches (e.g., original query = \query{http vs grpc}; expanded query = \query{http vs grpc performance}; Google Related Searches = \{\query{grpc vs rest performance},\query{grpc vs protobuf},...\}). While this indicates that most of the time (79\%) the developer's queries formulated by using our clarifying questions were different, many queries provided by Google Related Searches were topically related to the queries expanded by the clarification questions. However, we observed that in 12/42 (29\%) queries, all of the 8 queries in Google Related Searches were completely and significantly different from the expanded queries generated by the developers (e.g., original query = \query{unity mrtk detect pointer in collider}; expanded query = \query{unity mrtk detect pointer in collider fix tutorial}; Google Related Searches = \{\query{mrtk gestures},\query{mrtk disable pointer},...\}). Our hypothesis is that this scenario occurs when the original queries are longer and more specific as in that case Google Related Searches suggest more simple and generic alternatives that significantly miss the topical direction that the developer is interested in.}

\subsection{Threats to Validity}
Several limitations may impact the interpretation of our findings on the proposed approach for automatically posing clarification questions to help developers' Web search. We categorize and list each of them below.

\noindent {\bf Construct validity.} Threats to construct validity include the degree to which our experiments measure what was intended. First, as our original Web query dataset includes only 200 queries from a relatively narrow scope (i.e., Java software development), the model's training dataset is limited by the topics within those queries as well. More specifically, the validity concern related to this is whether the clarification questions and the model for predicting clarification questions overfit a specific set of topics. One mitigating factor is augmenting our dataset (using RoBERTa) by replacing query terms, which actually generates queries that are no longer Java-specific (e.g., replacing the term \query{java} with the term \query{python}). Another means to mitigate this concern is the two ways of evaluation of our model and technique, both indicating that the clarification questions and the model were generally relevant to the developer's search tasks. For instance, in the developer survey, 77/103 (75\%) of clarification questions were deemed relevant, which is very similar to the model's quantitative evaluation results for Precision@1 of 0.8 (80\%). 

Another threat is the use of common answers as input to our model. As the common answers are based on facets related to queries, they may be too specific to the particular domain, and, again, lead to overfitting. With different and more diverse common answers, the model may learn different patterns. We mitigate this issue by selecting the broadest common terms, which are likely to be valid across many different software engineering-specific topics.

\noindent {\bf Internal validity.} Threats to internal validity concern study design factors that may influence the results. One threat to our study is the artificially generated dataset. A mitigating factor is the use of a powerful natural language model, i.e., RoBERTa, to suggest terms to augment our queries. However, it is possible that some of the queries may be unrealistic and never occur word-for-word in real-life. We address this threat by manually checking each augmented query to be semantically correct and to still be in need of a specific clarification question.

\noindent {\bf External validity.} Threats to external validity concern generalizing the experimental findings beyond our settings. One such limitation of our approach is the fixed number of clarification questions, which are also limited by our input set of 200 Java-related Bing queries. It is possible to elicit additional clarification questions with a larger dataset of queries. For example, we observe that the common clarification question `\textit{Which programming language are you interested in?}' is missing in our study. This is because our annotated query dataset contains only Java-related queries and each query explicitly mentions `java'. To mitigate this threat, we performed a tool-based study with developers that worked on other development tasks that did not necessarily include Java (e.g., Go, Unity, C\#, PHP). 

\section{Related Work} \label{related_work}
The related work can be categorized into three parts: 1) empirical studies of software development-related Web search, 2) generating facets from queries, and 3) automatically posing clarification questions.

\noindent
{\bf Empirical Studies of Software Development-Related Web\\ Search.} Developers use general purpose search engines daily to facilitate their work. Therefore, there has been considerable interest among researchers in studying how developers perform these searches~\cite{brandt2009two, sadowski2015developers, xia2017developers, martie2017understanding, rahman2018evaluating, hucka2018software, Rao2020AnalyzingWS, sim2011well, hora2021googling}. There have been several studies on improving search specifically for source code~\cite{bajracharya2006sourcerer, martie2017understanding, kim2018facoy, holmes2009developers}. For example, Stolee et al.'s survey notes that 85\% of the software developers use search engines to search for source codes on the Web at least weekly~\cite{stolee2014solving}. They also found that 69\% of the respondents used Web search engines (i.e., Google, Bing) instead of bespoke code search tools. Xia et al. noted that developers search for many things other than source code, e.g., debugging, documentation, best practices, how to use a particular tool, etc.~\cite{xia2017developers} Temla et al. noted that game developers use Web search to get various kinds of help (e.g., instructions, algorithms, and tools)~\cite{tamla2019survey}. In addition, there has been research dedicated to developers' Web search for software architecture design patterns~\cite{soliman2021exploring}, algorithms~\cite{bhatia2011algorithm}, APIs~\cite{stylos2006mica}, etc.

Sadowski et al. surveyed 27 developers over a two week period to better understand when and why software developers search~\cite{sadowski2015developers}. Rahman et al. collected search logs from 310 developers, consisting of nearly 150,000 Google search queries and built an automatic classifier that detected software engineering related queries~\cite{rahman2018evaluating}. Hora et al. collected 1.3M developers' queries from a Search Engine Optimization service and concluded that the queries are usually short, tend to omit functional words, and are similar among each other~\cite{hora2021googling}. Xia et al. collected search queries from 60 developers and surveyed 235 software engineers and grouped search tasks into seven groups: General Search, Debugging and Bug Fixing, Programming, Third Party Code Reuse, Tools, Database, and Testing~\cite{xia2017developers}. Rao et al. used search query logs from Bing to analyze Web search behavior for software engineering tasks and classified the search intents into seven categories: API, Debug, HowTo, Installation, Learn, Navigational and Miscellaneous~\cite{Rao2020AnalyzingWS}. 

However, most of the above mentioned datasets are not publicly available to analyze further. In an attempt to encourage contributions to this area, Rao et al. open-sourced a portion of their Bing dataset~\cite{rao2020code}.

\noindent
{\bf Generating Facets from Queries.} Manually and automatically extracting facets from Web query logs has been an area of interest in the research community for a long period of time due to its potential for query reformulation~\cite{stoica2007automating, dou2011finding, dou2015automatically, kong2013extracting, kong2014extending, aliannejadi2019asking, kong2016precision, friedrich2015utilizing}. Li et al. proposed a system that can automatically generate facets for a query using Wikipedia~\cite{li2010facetedpedia}. Dou et al. developed an unsupervised model that automatically extracts facets for a query~\cite{dou2011finding, dou2015automatically}.  Kong et al. developed a supervised graphical model to extract facets for a query based on a set of Web search results~\cite{kong2013extracting, kong2014extending}. Jiang et al. proposed a model to generate facets from knowledge bases~\cite{jiang2016generating}. Aliannejadi et al. applied a manual facet generation process similar to ours for a corpus of queries~\cite{aliannejadi2019asking}. Specifically aimed at software engineering-related search, Zhang et al. used Stack Overflow tags as facets~\cite{zhang2020chatbot4qr}. 

\noindent
{\bf Automatically Posing Clarification Questions.} 
Automatically posing clarification questions has recently become an popular area of study in the fields of Natural Language Processing and Information Retrieval~\cite{aliannejadi2019asking, zhang2018towards, zhang2020chatbot4qr, ren2020conversations, rosset2020leading, hien2020towards, elgohary2019can, vakulenko2020question, anantha2020open, stoyanchev2014towards, zamani2020generating}. The research on asking clarification questions has been applied within a few different domains and applications such as chatbots~\cite{hancock2019learning}, open-domain question answering systems~\cite{de2003analysis, de2005implementing}, domain-specific question answering systems~\cite{zhang2020chatbot4qr, trienes2019identifying, braslavski2017you}, search engines~\cite{ren2020conversations}, \updatetext{search queries~\cite{wang2021template}}, image content~\cite{mostafazadeh2016generating}, and e-commerce~\cite{zhang2018towards}.

Clarification questions are a mechanism to refine user Web search queries that miss key information. Aliannejadi et al. built a model for query refinement to aid open-domain information seeking conversations~\cite{aliannejadi2019asking}. Zamani et al. proposed a template-based clarification question generation model based on different aspects of a Web query~\cite{zamani2020generating}. Zamani et al. later open-sourced a dataset for open domain clarification question generation~\cite{zamani2020mimics} and provided an analysis of user interaction with clarification questions posed in a popular Web search engine (Bing)~\cite{zamani2020analyzing}. Sekuli{\'c} et al. studied how to predict user engagement with posed clarification questions based on the open source Web search dataset released by Zamani et al.~\cite{sekulic2021user} In our work, we focus on software development-related clarification questions and release a dataset of annotated queries, facets, and clarification questions.

\section{Conclusions and Future Work}
This paper illustrates a technique for posing clarification questions for developers' Web search queries. In this paper, first, we identify common clarification questions based on annotating 200 real-world user queries. Then, we create an augmented dataset of 5,151 queries that we use to train a machine learning model that can identify the most relevant clarification question for a given user query. Our evaluation shows that the proposed approach outperforms all baselines, while a study of software developers indicates that the clarification question relevance rate aligned with our quantitative (i.e., test set) results. That is, the participants found 75\% clarification questions were relevant, while the Precision@1 on the test set was 80\%. The feedback from the software developer study, using a prototype we created, suggests that an approach like ours can be useful to software developers.

There are several avenues for future work. First, we intend on making the common answers dynamic, i.e., more adaptable to the specific query. Second, we plan to research an interactive conversation based tool where the users can ask clarification questions successively and engage the search system in a type of a conversation. Third, another line of future work is in identifying different set of clarification questions based on query intents proposed by various literature~\cite{xia2017developers, Rao2020AnalyzingWS}. Lastly, we can extend the evaluation to a larger set of developers, which would help in further understanding how to improve the current model and clarification questions.

\bibliographystyle{elsarticle-num}
\bibliography{paper}

\end{document}